\documentstyle[11pt]{article}
\begin{document}

\begin{titlepage}
\hfill MS-TPI-94-2 \\

\begin{centering}
\vfill
{\LARGE \bf Algebraic Computation of Hierarchical \\
            Renormalization Group Fixed Points \\[2mm]
            and their $\epsilon$-Expansions}

\vspace{2cm}
{\bf K. Pinn, A. Pordt, and C. Wieczerkowski} \\[6mm]

\vspace{0.3cm}
{\em
\, Institut f\"ur Theoretische Physik I, Universit\"at M\"unster, \\
   Wilhelm-Klemm-Str.\ 9, D-48149 M\"unster, Germany \\}

\vspace{2cm}
{\bf Abstract} \\
\end{centering}
\vspace{0.2cm}

Nontrivial fixed points of the hierarchical renormalization group  are
computed by numerically solving a system of quadratic   equations for
the coupling constants. This approach avoids a fine tuning of relevant
parameters. We study the eigenvalues of the renormalization group
transformation, linearized around the non-trivial fixed points.  The
numerical results are compared with $\epsilon$-expansion.

\vspace{0.3cm}\noindent

\vfill \vfill
\noindent
MS-TPI-94-2\\
March 1994
\end{titlepage}
\newcommand{\nc}{\newcommand}
\nc{\be}{\begin{equation}}
\nc{\ee}{\end{equation}}
\nc{\bea}{\begin{eqnarray}}
\nc{\eea}{\end{eqnarray}}
\nc{\rbo}{\raisebox}
\nc{\cH}{{\cal H}}
\nc{\RR} {\rangle \! \rangle}
\nc{\LL} {\langle \! \langle}
\nc{\rmi}[1]{{\mbox{\small #1}}}
\nc{\eq}{eq.~}
\nc{\nr}[1]{(\ref{#1})}
\nc{\ul}{\underline}
\nc{\cM}{{\cal M}}
\nc{\mc}{\multicolumn}
\nc{\todo}[1]{\par\noindent{\bf $\rightarrow$ #1}}

\nc{\nonu}{\nonumber}
\nc{\lmax}{l_{\rm max}}
\nc{\half}{\mbox{\small$\frac12$}}
\nc{\eights}{\mbox{\small$\frac18$}}
\nc{\bg}{\bar \gamma}


\newtheorem{lemma}{Lemma}[section]
\newtheorem{corollary}{Corollary}[section]
\newtheorem{theorem}{Theorem}[section]
\newtheorem{proposition}{Proposition}[section]
\newtheorem{definition}{Definition}[section]
\def\lst{{l^*}}
\def\dst{{d^*}}
\def\bR{{\bf R}}
\def\bZ{{\bf Z}}
\def\cC{{\cal C}}
\def\cN{{\cal N}}
\def\cV{{\cal V}}
\def\om{{\overline m}}
\def\g{{\gamma}}
\def\l{{\lambda}}
\def\L{{\Lambda}}
\def\rv{{\rm v}}
\def\Gau#1#2{d\mu_{#1}({#2})}
\def\iGaurv{\int d\mu_{\rv }({\phi })}
\def\inn{{\underline \in }}
\def\xp{{x^{\prime }}}
\def\zp{{z^{\prime }}}

\thispagestyle{empty}
\newpage

\section{Introduction}
\label{SECintro}

The renormalization group (RG) \cite{RG} is a nonperturbative approach to
critical phenomena in statistical mechanics and Euclidean field theory.
It offers a satisfactory explanation for the behavior of statistical
models in the critical regime, in particular the appearance of universal
quanities. It also suggests block spin transformations as an actual
scheme for computations of critical quantities. Last not least it has
proved to be a powerful tool in rigorous investigations. Although a lot
of efforts have been put into the subject since the seminal work of
Wilson it is far from being closed.

The main idea is to think of a critical system in terms of the flow of
effective actions generated by a block spin transformation. This flow
will have fixed points. An instructive example is a system with two
possible phases separated by a second order transition in the absense of
an external field. Then we typically find two stable fixed points
characterizing the different phases and one unstable fixed point
signalling a phase transition. Critical indices can be calculated as
eigenvalues of a linearized transformation at the unstable fixed point.
Given any critical system the most important task is thus the
determination of its RG flow fixed points and the
flow in their vicinity.

Let us consider a Euclidean scalar field on a $d$-dimensional unit
lattice $\Lambda$. The action is $S(\phi)=S_0(\phi)+V(\phi)$ with
$V(-\phi)=V(\phi)$. The interaction is called local if the Boltzmann
factor ${\rm exp}(-V(\phi))$ factorizes into a product
$\prod_{x\in\Lambda} F(\phi(x))$. Unfortunately, locality is only
preserved in an approximate sense by block spin transformations. A main
technical difficulty therefore is to find an efficient parametrization
of the effective actions generated by the flow. Also it is not a priori
clear what the best choice for the block spin is. A guideline is to
demand the effective actions to be approximately as local as possible.

The block spin transformation which we will consider here relies on a
decomposition of the Gaussian measure $d\mu_\rv (\phi)$ determined by
the free action $S_0(\phi)=\frac12 (\phi,\rv^{-1}\phi)$ into a
background and a fluctuation part. The free covariance (propagator) is
$\rv=(-\Delta)^{-1}$. The block spin transformation then consists of
an integration step, a rescaling step of the background field and the
block lattice, and a subsequent subtraction of the field independent
term. In this setup the non-locality of the effective interaction is
traced back to the fact that the fluctuation covariance despite its
exponential decay couples fields further apart than a block distance.

In the hierarchical model \cite{hiera}  the free covariance
$(-\Delta)^{-1}$ is replaced by a hierarchical counterpart
$(-\Delta)_{{\rm hier}}^{-1}$. Both covariances share a similar long
distance decay, although the hierarchical covariance is not translation
invariant, and the models therefore behave similarly in the critical
region. As a gain the hierarchical block spin transformation preserves
locality in a strict sense. A hierarchical RG step
turns out to be a nonlinear transformation of the local Boltzmann factor
$F(\phi)$, which is a function of a single variable rather than a
functional of a field configuration.

One point of view is to think of a hierarchical model as an
approximation to its full brother. This is obviously only justified if
there exists an interpolation between the two which admits the
computation of systematic corrections around the hierarchical situation.
Such a scheme is still lacking although a number of first steps into
this direction can be found in the literature, see, e.g., \cite{golner}.
The main difference
between the hierarchical and the full model is the absence of wave
function renormalization in the former. Wave function renormalization
comes about in the full model because the effective interaction contains
a kinetic term as a part of the quadratic piece. This is
clearly discarded once the effective action is forced to be ultra-local.

A second point of view is to think of hierarchical models as nontrivial
statistical models by themselves which are perfectly suited for a
RG approach. Due to their nontrivial interaction
they are far from being exactly solvable. Their phase structure and
critical behavior is as rich and complicated as in the full case. This
point of view seems somewhat academic since we do not know of any real
ensembles which belong to the universality class of a hierarchical
model. Real systems do not show this peculiar breaking of translation
invariance.

The third point of view is to consider hierarchical models as a test
ground for both concepts and methods towards a RG
treatment of the full model. The problem of non-localities in the
effective action is believed to be of purely technical nature. This is
convincingly supported by rigorous studies of many full critical models
\cite{rigor} which have been preceeded by studies of their hierarchical
counterparts. A full RG flow will always contain as a
part a flow of local quantities which is accurately modelled by a
hierarchical flow.

An exciting possibility within the hierarchical context is to tune the
dimension parameter $d$ away from integer values. The hierarchical
transformation does not involve any lattice geometry and contains $d$
and a scale parameter $L$ as variables which can be taken real valued.
Let us stress that this can be done completely independent of
perturbation theory. A beautiful picture emerges that shows a sequence
of thresholds between two and four dimensions where new fixed points
appear. For instance, four dimensions is the threshold below which an
unstable double well fixed point exists which governs three dimensional
physics. This picture is believed to be true also in the full model
although we do not know of a RG formulation which
incorporates lattice geometry only in form of a dimension parameter
which can be tuned at will.

In this paper we investigate the question of RG flows
in the hierarchical model. Usually small coupling constants are required
to do analytically  such an investigation. To study infrared fixed
points we have no small  coupling parameters at hand. Nevertheless, we
will show that we  can do RG calculations for models which are not
asymptotically free by using some algebraic equations.

After splitting off the quadratic fixed point, which corresponds to the
unordered phase, we expand the Boltzmann factor in terms of local
operators. We will use two kinds of operators, simple powers of fields
and normal ordered products. Then the fixed point equation becomes a set
of algebraic equations for the coupling constants. We truncate this
system and solve it numerically. This turns out to be much better
than searching for fixed points by iterating block spin transformations
and fine tuning the initial values. It allows us to determine even the
higher fixed points with multi-dimensional unstable manifolds to high
accuracy. Then we linearize this set of equations and diagonalize it
numerically around the various fixed points. This directly computes the
critical indices of the hierarchical model without having to rely on any
kind of expansions.

The algebraic equations can also be written down for the full model in
terms of a polymer expansion. It is less clear then how to truncate this
system to a reasonable size. But we believe that this is managable and
will admit a similar numerical treatment.

The algebraic equations can be solved for any value of $d>2$.
Inserting a power series ansatz in $\epsilon$ for the coupling
constants at $d=4-\epsilon$, we obtain the $\epsilon$-expansion for the
double well fixed point and analogously at the other critical
dimensions. The resulting set of equations can be put into a recursive
form which we iterate exactly using computer algebra. We compare this
$\epsilon$-expansion with the numerical results of the previous
computations both for the fixed points and the critical indices.

Finally, we investigate the flow around the fixed points in terms of the
eigenoperators of the linearized flow. It turns out that the
eigenoperators at the double well fixed point are approximately normal
ordered powers in a small field region but show exponential decay at
large fields.

Hierarchical models have been investigated by many authors. Closest to
our approach is the constructive work of Koch and Wittwer on the
nontrivial fixed point \cite{KochWittwer}. The bifurcation picture is
also described within the framework of a continuous hierarchical
RG (as opposed to the lattice approach) in the
article of Felder \cite{Felder}. Another investigation of the
$\epsilon$-expansion for the double well fixed point was done by Collet
and Eckmann \cite{ColletEckmannEps}.

The present contribution to the subject is an investigation of the flow
on the computer in terms of an algebraic formulation both for the fixed
points, the critical indices, the eigenoperators, and the
$\epsilon$-expansion.

\section{The Model}
\label{SECmodel}

We consider hierarchical RG transformations, defined by
\bea\label{trafofull}
 F \mapsto F', \qquad
 F'(\phi) &=& \int d\mu_{\gamma}(\zeta) \,
 F(\zeta + \beta \phi)^{L^d} \, ,
 \nonu \\
 \beta  &=& L^{1-d/2} \, .
\eea
In Appendix 1, we  discuss the hierarchical model as an approximation of
a field theoretic model and motivate the hierarchical RG
transformation.

$d\mu_{\gamma}$ denotes the 1-dimensional Gaussian
measure with covariance $\gamma > 0$, defined through
\be
 \int d\mu_{\gamma}(x) f(x) = (2\pi \gamma)^{-1/2}
 \int_{-\infty}^{+\infty} dx \,
 \exp\left(- \frac{x^2}{2\gamma}\right) \, f(x) \, .
\ee
$L$ denotes the block size of the corresponding blocking transformation.
It is not difficult to convince oneself that
\be
 F_{HT}(\phi) = L^{1/(L^d-1)} \cdot \exp \left(
 - \frac{L^2-1}{2 \gamma L^d} \, \phi^2 \right)
\ee
is a fixed point of the transformation eq.\ (\ref{trafofull}).
$F_{HT}$ is called high temperature fixed point.
Let us write the Boltzmannian $F$ as a product,
\be\label{splitit}
 F(\phi) = F_{HT}(\phi) \, Z(\phi) \, .
\ee
The RG transformation of $Z$
is again of the type eq.\ (\ref{trafofull}),
\be\label{trafosplit}
 Z \mapsto Z', \qquad
 Z'(\phi) = \int d\mu_{\gamma'}(\zeta) \, Z(\zeta + \beta' \phi)^{L^d} \, .
\ee
with the changed parameters $\gamma'$ and $\beta'$ given by
\be
 \beta'  =  L^{-1-d/2} \, , \qquad
 \gamma' =  L^{-2} \gamma \, .
\ee

In the following, we shall always choose
$L^d=2$. As we shall see below, this choice will
lead to quadratic fixed point equations. Note, however, that
it is not obvious how to realize a block spin transformation
for such a scale factor on a lattice.
For $L^d=2$, we have
\be
 \beta  = 2^{-(d-2)/(2d)} \, , \qquad
 \beta' = 2^{-(d+2)/(2d)} \, .
\ee
Furthermore, we choose $\gamma$ such that $\gamma'=\half(1-{\beta'}^2)$, i.e.,
\be\label{gamma}
 \gamma  = \half \left( 2^{2/d} - \half   \right) \, .
\ee
This type of conventions agrees with those of ref.\ \cite{KochWittwer}.
We study in this paper the fixed points of the RG transformation
eq.\ (\ref{trafosplit}) that are even in $\phi$,
i.e., the even solutions of the fixed point equation
\be\label{fixpo}
Z(\phi) = \int d\mu_{\gamma'}(\zeta) \, Z(\zeta + \beta' \phi)^{L^d} \, .
\ee
Let us remark that if $Z$ is a solution of the fixed point equation
(\ref{fixpo}), then $Z_\delta $, where $Z_\delta (\phi ):=
Z(\delta \phi )$, $\delta >0$, is a solution of eq.~(\ref{fixpo})
if $\gamma' $ is replaced by $\gamma' \delta^{-2}$.
The consequence of dealing with $Z$ instead of $F$ is
that $\beta$ is replaced by $\beta' < \beta$. This has
important consequences in the constructive approach of
Koch and Wittwer \cite{KochWittwer}.
\section{Computation of Nontrivial RG Fixed Points}
\label{SECfix}

A necessary ingredient for a successful study of the transformation
(\ref{trafofull}) is a good choice of coordinates.
Good here means that already a reasonable number of coupling
constants provide a good approximation of the full problem.
Furthermore, good coordinates should be also easy to deal  with in
practical calculations.

To study the transformation (\ref{trafofull}), one might  consider the
expansion of $F(\phi)$ in powers of $\phi$.  However, if such an
expansion is truncated at a finite order,  the resulting approximation
will probably not define a  reasonable (i.e., positive) Boltzmannian.
The problem is that the coefficients change their sign  from order to
order, like in the Taylor expansion of  $\exp(-x^2)$.

A good choice (though perhaps not the best) is to
use the split eq.~(\ref{splitit}), and consider a Taylor expansion
of $Z(\phi)$ in powers of $\phi$. This choice of coordinates
has been successfully employed in a rigorous proof that the
transformation (\ref{trafofull}) has a nontrivial fixed point
in $d=3$ dimensions \cite{KochWittwer}.
It has the advantage that
the expansion coefficients do not fluctuate in sign. Furthermore,
the correct large $\phi$ behavior of the Boltzmannian
is implemented automatically.

\subsection{Expansion of $Z$ in Powers of $\phi^2$}

Let us define a rescaled function $P(\phi)$ through
\be
  Z(\phi) = P \left( \phi / \sqrt{2 \gamma'} \right) \, .
\ee
The RG transformation for $P$ is
\be\label{ptrafo}
 P \mapsto P', \qquad
 P'(\phi) = \int d\mu_{1/2}(\zeta) \, P(\zeta + \beta' \phi)^2 \, .
\ee
We expand $P(\phi)$ in powers of $\phi^2$,
\be\label{simple}
 P(\phi) = \sum_{l \geq 0}
 p_l \, \frac{ \phi^{2l} }{ 2^l \, \sqrt{(2l)!}  } \, .
\ee
The specific choice of normalization of the $p_l$ in eq.\ (\ref{simple})
turned out to be suitable for the numerical fixed point solver
to be described below.
In terms of the expansion coefficients $p_l$, the RG reads
\be
  p_l \mapsto  p_l' = \sum_{m,n} S_l^{mn} \, p_m p_n \, ,
\ee
with
\be
 S_l^{mn} = \left\{ \begin{array}{ll}
 \beta'^{2l} \, (\eights)^{-l+m+n}
 \frac{(2(m+n))!}{(m+n-l)! \sqrt{(2l)!(2m)!(2n)!}}
 &\mbox{if} \mbox{\small $\quad 0 \leq l \leq m+n$} \, , \\
 0  & \mbox{else} \, .
 \end{array}
 \right.
\ee
If we look for a RG fixed point, we have to study the
infinite set of quadratic equations
\be\label{fixeq}
  0 = p_l^* - \sum_{m,n} S_l^{mn} \, p_m^* p_n^* \, .
\ee

\subsection{Numerical Solution of the Fixed Point Equations}

A straightforward numerical treatment of the problem defined  by
eq.~(\ref{fixeq}) becomes possible  if we truncate the sum over $l$ in
eq.~(\ref{simple}), introducing a highest index $\lmax$,
\be\label{trunc}
 P(\phi) = \sum_{l = 0}^{\lmax}
 p_l \, \frac{ \phi^{2l} }{2^l \, \sqrt{(2l)!}  } \, .
\ee
Then, of course, also the sums over $m$ and $n$ run only from
$0$ to $\lmax$, and the fixed point problem consists in the
study of $\lmax+1$ quadratic equations,
\bea\label{fixtrunc}
  f_l &=& 0 \, , \quad  0 \leq l \leq \lmax \, , \nonu \\
  f_l &=& p_l - \sum_{m,n \leq \lmax} S_l^{mn} \, p_m p_n \, .
\eea
We shall denote solutions of these equations by $p_l^*$.
We used the routine C05NBF from the NAGLIB library
\cite{NAG} for a numerical solution of the fixed point equations in the
range $2 < d < 4$. Notice that the dimension dependence enters through
the $\beta'$-dependence of the structure coefficients $S_l^{mn}$.
The program requires an initial guess of the solution. If the program
is successful, it returns a solution, together with the values
of the $f_l$ for this particular solution.
In table \ref{demotab}, we show as an example the $p_l^*$ for
$\lmax=10$ and $d=3$, together with the $f_l^*$ that can be
considered as a measure of the error of the solution.

\begin{table}
\small
\begin{center}
\begin{tabular}{|r|c|r|}
\hline
  \mc{1}{|c|}{$l$}  &
  \mc{1}{c|}{$p_l^*$} &
  \mc{1}{c|}{$f_l$} \\
\hline
  0 & $0.752806717034\cdot 10^{+0}$ & $  0.423398205312\cdot 10^{-16}$ \\
  1 & $0.481272697982\cdot 10^{+0}$ & $ -0.113124959171\cdot 10^{-17}$ \\
  2 & $0.313506765870\cdot 10^{+0}$ & $  0.347242613663\cdot 10^{-17}$ \\
  3 & $0.186261032043\cdot 10^{+0}$ & $ -0.179714118083\cdot 10^{-17}$ \\
  4 & $0.100696164171\cdot 10^{+0}$ & $  0.395226286632\cdot 10^{-17}$ \\
  5 & $0.499270725225\cdot 10^{-1}$ & $  0.117550777471\cdot 10^{-16}$ \\
  6 & $0.228929876623\cdot 10^{-1}$ & $  0.570517883570\cdot 10^{-17}$ \\
  7 & $0.977563729148\cdot 10^{-2}$ & $  0.622745240266\cdot 10^{-17}$ \\
  8 & $0.390718140134\cdot 10^{-2}$ & $  0.471372511658\cdot 10^{-17}$ \\
  9 & $0.146546809430\cdot 10^{-2}$ & $  0.307351198867\cdot 10^{-17}$ \\
 10 & $0.515497714660\cdot 10^{-3}$ & $  0.122533903732\cdot 10^{-17}$ \\
\hline
 \end{tabular}
  \parbox[t]{.85\textwidth}
  {
  \caption[$p_l^*$ and $f_l^*$ for $\lmax=10$]{\label{demotab}
  The $p_l^*$ and $f_l$ for $\lmax=10$ in $d=3$ dimensions}
  }
\end{center}
\end{table}
We used truncation parameters $\lmax$ in the range $10 \dots 50$.
For the 2-well fixed points, $\lmax=20$ was completely sufficient.
To get an impression of the finite $\lmax$ effects, the reader
is invited to study table \ref{tabfin}, where we give the
first 11 coefficients of the 2-well fixed point in three dimensions
for $\lmax=10$, $20$, and $30$.

\begin{table}
\nc{\Aa}{\cdot 10^{-1}}
\nc{\Bb}{\cdot 10^{-2}}
\nc{\Cc}{\cdot 10^{-3}}
\nc{\Dd}{\cdot 10^{-0}}
\small
\begin{center}
\begin{tabular}{|r|l|l|l|}
\hline
  \mc{1}{|c|}{$l$}  &
  \mc{1}{c|}{$\lmax=10$} &
  \mc{1}{c|}{$\lmax=20$} &
  \mc{1}{c|}{$\lmax=30$} \\
\hline
  0 & $0.752806717034\Dd$ & $0.752859732932\Dd$ & $0.752859732933\Dd$ \\
  1 & $0.481272697982\Dd$ & $0.481191004612\Dd$ & $0.481191004610\Dd$ \\
  2 & $0.313506765870\Dd$ & $0.313445974082\Dd$ & $0.313445974081\Dd$ \\
  3 & $0.186261032043\Dd$ & $0.186254920100\Dd$ & $0.186254920100\Dd$ \\
  4 & $0.100696164171\Dd$ & $0.100729195010\Dd$ & $0.100729195011\Dd$ \\
  5 & $0.499270725225\Aa$ & $0.499755196476\Aa$ & $0.499755196485\Aa$ \\
  6 & $0.228929876623\Aa$ & $0.229416956212\Aa$ & $0.229416956222\Aa$ \\
  7 & $0.977563729148\Bb$ & $0.981866078064\Bb$ & $0.981866078150\Bb$ \\
  8 & $0.390718140134\Bb$ & $0.394316100256\Bb$ & $0.394316100333\Bb$ \\
  9 & $0.146546809430\Bb$ & $0.149410716099\Bb$ & $0.149410716173\Bb$ \\
 10 & $0.515497714660\Cc$ & $0.536648340900\Cc$ & $0.536648341727\Cc$ \\
\hline
 \end{tabular}
  \parbox[t]{.85\textwidth}
  {
  \caption[Finite $\lmax$ effects]{\label{tabfin}
  The first 11 coefficients $p_l^*$ of the 2-well in $d=3$ dimensions
  for three different values of $\lmax$}
  }
\end{center}
\end{table}
We define an effective {\em potential} $V(\phi)$ through
\be
V(\phi) := -\ln \left( \frac{F(\phi )}{F(\phi=0)} \right) \, .
\ee
In figure 1, we show our results for the 2-well potential
for $d=2.1$ through $d=3.8$ in steps of $0.1$.
In all cases we used $\lmax=20$. The deepest potential
corresponds to $d=2.1$. With increasing dimension, the 2-well
gets flatter and flatter until it vanishes in four dimensions.

{}From naive power counting and the studies of section \ref{SECeps},
one expects that $n$-well fixed points
occur when the dimension goes below the threshold
$d_n = 2n/(n-1)$. Note that exactly at these thresholds the
operators $\phi^{2n}$ become relevant with respect to the
Gaussian fixed point. Thus 2-wells are expected to exist
for $d<4$, 3-wells for $d<3$, 4-wells for $d<8/3$ and so on.

By starting our program with a suitable initial guess, we
were able to find the 3-well potentials. Figure 2 shows
the results for $d=2.1$ through $d=2.7$ in steps of $0.1$.
The deepest potential corresponds to $d=2.1$, the flattest
one to $d=2.7$. In all cases we used again $\lmax=20$.

The search for the 4-wells was a bit more difficult. It turned
out that one needed more than 20 couplings for a reasonable
parameterization. Figure 3 shows the 4-well potential in
$d=2.1$ and in $d=2.3$ dimensions. We used $\lmax=30$ in
both cases.

In all cases, we made a further check on our results as follows:  We
represented the Boltzmannian $F$ by its values on a grid  of typically
300 sites on the interval $0 \leq \phi \leq \Lambda$,  where $\Lambda$
was typically 10 or 20. The RG steps were then done by numerically
performing the integral \ref{trafofull},
using the NAGLIB routine D01GAF. This routine
determines an approximation of an integral if the integrand is given on
a finite number of points.  All fixed points determined by the algebraic
method were converted to functions on the $\phi$-grid and then checked
for stability under iterated application of the integration  method.

\section{The Linearized Renormalization Group}
\label{SECexpo}

In this section we shall report on a numerical study of the  eigenvalues
of the hierarchical RG, linearized around the  nontrivial fixed points.
Note that the eigenvalues related to the full transformation
(\ref{trafofull}) are the same  as those related to the transformation
(\ref{trafosplit}) for $Z$. The reason is that fixed points $F^{*}$ and
$Z^{*}$ differ only  by the factor $F_{HT}$ that is a fixed point
itself.  The transformation for $Z$ around a fixed point  parameterized
by coordinates $p_l^{*}$ is
\be
  (p_l^{*} + \delta_l) \mapsto  (p_l^{*} + \delta_l')
  = \sum_{m,n} S_l^{mn} \, (p_m^{*} + \delta_m) \, (p_n^{*} + \delta_n) \, .
\ee
We expand to first order in $\delta$,
\be
  \delta_l \mapsto  \delta_l'
  = \sum_{n} R_{ln} \, \delta_n \, ,
\ee
and identify the linearized RG transformation with the matrix $R$,
\be
  R_{ln} = 2 \sum_m \, S_l^{mn} \, p_m^{*} .
\ee
The matrix $R$ is not symmetric. It can, however, be shown that its
eigenvalues are real, cf.\ lemma \ref{lemma3.1} below.
We used the NAGLIB procedure F01AKF and F02APF to compute
all eigenvalues of $\lambda_i$ of $R$, with $i=0,\dots,\lmax$.
It turns out that all eigenvalues are positive and can
therefore be exponentiated.
We define
\be
 \lambda_i = L^{a_i} \, ,
\ee
where $L$ denotes the block size, in our case $2^{1/d}$.

Table \ref{compexp} shows, as an example, the first five eigenvalues
$a_i$ of the linearized RG around the 2-well, the 3-well, and the 4-well
in $d=2.1$ dimensions, respectively.  Positive eigenvalues $a_i$ are called
relevant, negative eigenvalues  are called irrelevant, and a zero
eigenvalue is called marginal.  In accordance with the expectation, the
$n$-wells have $n-1$  (nontrivial) relevant eigenvalues.

\begin{table}
\small
\begin{center}
\begin{tabular}{|c|r|r|r|}
\hline
  $i$ &
 \mc{1}{|c|}{2-well}  &
 \mc{1}{c|}{3-well} &
 \mc{1}{c|}{4-well} \\
\hline
 0 &  2.1000000 &  2.1000000 &  2.1000000 \\
 1 &  0.4787297 &  1.9715290 &  1.9851950 \\
 2 & -1.1724335 &  0.4611953 &  1.0966180 \\
 3 & -3.3445644 & -0.7546772 &  0.4105210 \\
 4 & -5.8303542 & -2.2270214 & -5.6565547 \\
 5 & -8.4946711 & -3.9790124 & -1.7706071 \\
\hline
 \end{tabular}
  \parbox[t]{.85\textwidth}
  {
  \caption[The 6 leading eigenvalues at $d=2.1$]{ \label{compexp}
  The 6 leading eigenvalues $a_i$ at $d=2.1$ for the
  2-well, the 3-well, and the 4-well, respectively}
  }
\end{center}
\end{table}
One always observes an eigenvalue $a_0 = d$ that corresponds
to the trivial volume operator. It corresponds to the fact that the fixed
point $Z^*$ itself is an eigenvector with eigenvalue $\lambda_0=2$.
The next eigenvalue $a_1$ is
related to the critical exponent $\nu$, via \cite{Bellac}
\be
\nu = \frac{1}{a_1} \, .
\ee
In table \ref{tabnus} we show our results for the exponent  $\nu$ for
$d=2.1$ through $d=3.8$ in the case of a 2-well.  It is interesting to
compare the $d=3$ result with the exponent $\nu$ for  the 3-dimensional
Ising model. The best known estimates for the latter are in the range
$0.624 \dots 0.630$ \cite{fulnu}. For $d \rightarrow 2$, the deviation
of the exponent  $\nu$ from the value in the full model increases. We
know from the  exact solution of the 2-dimensional Ising model that
$\nu=1$ for $d=2$. In the hierarchical model we observe $\nu > 2$ already
at $d=2.1$.
\begin{table}
\small
\begin{center}
\begin{tabular}{|c|c||c|c|}
\hline
  $d$ & $\nu$  & $d$ & $\nu$ \\
\hline
2.1 & 2.08886 & 3.0 & 0.64957 \\
2.2 & 1.36234 & 3.1 & 0.62570 \\
2.3 & 1.09916 & 3.2 & 0.60484 \\
2.4 & 0.95704 & 3.3 & 0.58640 \\
2.5 & 0.86534 & 3.4 & 0.56995 \\
2.6 & 0.79985 & 3.5 & 0.55516 \\
2.7 & 0.74993 & 3.6 & 0.54182 \\
2.8 & 0.71011 & 3.7 & 0.52973 \\
2.9 & 0.67729 & 3.8 & 0.51877 \\
\hline
 \end{tabular}
  \parbox[t]{.85\textwidth}
  {
  \caption[Results for the critical exponent $\nu$]{\label{tabnus}
  Results for the critical exponent $\nu$ in the hierarchical model
  for the 2-well}
  }
\end{center}
\end{table}

It is interesting to look also at the eigenvectors of the
linearized RG transformation.
As an example we consider the eigenvectors of the transformation
eq.\ (\ref{trafofull}), linearized around the trivial fixed point
$F^*(\phi)=1$ and around the nontrivial 2-well fixed point
in $d=3$ dimensions. The eigenvectors $O_i^{(0)}(\phi)$ for
the linearization around $F^*=1$ can be given exactly:
\be
O_i^{(0)}(\phi)= \mbox{const} \cdot H_{2n} \left( \frac{\phi}{2\bar\gamma}
\right) \, ,
\ee
with
\be
  \bg \equiv \frac{\gamma}{1-\beta^2} \, .
\ee
The $H_k$ denote the Hermite polynomials.
We compare these functions for $i=1,2$ with the corresponding
eigenvectors of the RG linearized around the 2-well fixed point in
$d=3$, see figure 5.  The figure shows that the eigenvectors in $d=3$
are of similar  shape as the corresponding functions for the trivial
fixed point.  Note, however, that the ``full" eigenvectors contain a
factor $F_{HT}$ and thus have a completely different large $\phi$
behavior than the Hermite polynomials.

\section{$\epsilon$-Expansion}
\label{SECeps}

In this section, we want to compare the numerical results  obtained in
the previous section with expansions  in $\epsilon = \dst - d $, where
$\dst$ is one  of the threshold dimensions $d_n = 2n/(n-1)$.   This
section is organized as follows: In subsection \ref{epswick}  a
parametrization of the hierarchical Boltzmannian is introduced  that is
suitable for the $\epsilon$-expansion.  The recursion relations for the
expansion coefficients   $a_l^{(k)}$ shall be derived in subsection
\ref{epsal}. In subsection \ref{epslin} we  will derive the recursion
relations for the computation  of the $\epsilon $-expansion for the
eigenvalues and eigenvectors of the linearized $\epsilon $-expansion.
In subsection \ref{comparison} we shall give some comparisons
of the $\epsilon$-expansion results with the numerical results.

\subsection{Expansion of $F$ in Wick Monomials}
\label{epswick}

We consider the expansion of the  Boltzmannian $F$ in terms of normal
ordered powers of $\phi^2$.  This expansion shall be the basis for the
$\epsilon$-expansion to be studied below. One expands in terms of normal
ordered monomials
\be
F(\phi) = \sum_{l \geq 0} \frac{a_l}{\bg^l} \, : \phi^{2l} :_{\bg} \, .
\ee
The reason to divide the coefficients $a_l$ by $\bg^l$ is that
$a_l$ becomes $\bg^l$-independent for the fixed point.
Under a RG step the coordinates $a_l$ transform
according to
\be\label{newas}
a_l \mapsto a_l' = \beta^{2l} \sum_{m,n} {\cal C}_l^{mn} \, a_m a_n \, .
\ee
The sum in eq.\ (\ref{newas}) is restricted to $|m-n| \leq l \leq m+n$,
and the `structure coefficients' are given by
\be
{\cal C}_l^{mn} = \frac{(2m)! \, (2n)!}{(m+n-l)! \, (l+n-m)! \, (l+m-n)!} \, .
\ee
Normal ordering with respect to a covariance $\bg$ is defined through
\be
: \exp(a\phi) :_{\bg} = \exp\left( - \frac{\bg}{2} a^2 + a\phi \right) \, .
\ee
The normal ordered powers of $\phi$ can be expressed in terms of
Hermite polynomials,
\be
: \phi^n :_{\bg} = \left( \frac{\bg}{2}\right)^{n/2} \,
H_n \left( \frac{\phi}{\sqrt{2\bg}}   \right) \, .
\ee
That we use a direct expansion of $F$ as a basis for the
$\epsilon$-expansion seems to be in contradiction to our statement
above that it should be much better to use the split
$F=F_{HT} \, Z$, and then expand $Z$. The $\epsilon$-expansion
is, however, an expansion about $F=1$ (and {\em not} about $Z=1$).
Furthermore, as we shall see from the recursion relations to be
derived below, the effective Boltzmannians to a given order
in $\epsilon$ live in a finite dimensional space of coupling
constants. It is thus irrelevant which coordinates are chosen
in this space as long as they span this space.

\subsection{$\epsilon$-expansion for the $a_{l}$}
\label{epsal}

In this subsection we derive the recursive relations for the coefficients
$a_l^{(k)}$ defined through
\be
a_l = \sum_{k=0}^\infty a_l^{(k)} \epsilon^k \, .
\ee
We determine by $\epsilon $-expansion
the coefficients of the infrared fixed point at $d=\dst -\epsilon$
dimensions starting from the trivial fixed point in $\dst :=
2\lst /(\lst -1)$ dimensions, where $\lst \in \{ 2,3,\dots \} $.
The following lemma shows how to compute the $a_l^{(k)}$
recursively.
\begin{lemma}\label{lemma2.1}
Suppose that the coefficients $(\beta^{-2l})^{(m)}$ are defined by
\be
\beta^{-2l} = 2^{\frac{l}{\lst}} \sum_{m=0}^\infty (\beta^{-2l})^{(m)}
\epsilon^m \, .
\ee
For $\lst \in \{ 2,3,\ldots \} $ and $a_l^{(0)} = \delta_{l,0} \, $, we have
\be
a_l^{(1)} = \alpha \, \delta_{l,\lst } ,\qquad \alpha :=
            \frac{2(\beta^{-2\lst})^{(1)}}{\cC^{\lst \lst}_{\lst}} \, ,
\ee
and, for all $l \ne \lst $, $N\ge 1$,
\be
a_l^{(N)} = \frac{1}{2^{\frac{l}{\lst}}-2} \sum_{k=1}^{N-1} \left(
- 2^{\frac{l}{\lst}}
(\beta^{-2l})^{(N-k)} a_l^{(k)} + \sum_{m,n} \cC^{m n}_l a_m^{(N-k)}
a_n^{(k)} \right) \, .
\label{e2.2}
\ee
For all $N\ge 2$ \, ,
\bea
a_{\lst }^{(N)} = -\frac{1}{2(\beta^{-2\lst })^{(1)}} \biggl[
&-& 2 \biggl ((\beta^{-2\lst })^{(N)}\alpha
+ \sum_{k=2}^{N-1} (\beta^{-2\lst })^{(N+1-k)} a_{\lst}^{(k)}
\biggr ) + \nonumber \\
 &+& \sum_{k=2}^{N-1} \sum_{m,n} \cC^{m n}_{\lst} a_m^{(N+1-k)}
a_n^{(k)} +\nonumber\\
&+& 2\sum_{m:\ m\ne \lst} \cC^{m \lst}_{\lst} a_m^{(N)} \alpha \, \,
\biggr] \, .
\label{e2.3}
\eea
\end{lemma}

\vskip5mm\noindent
\underline{\sl Proof:} The $N$th order
term of the fixed point equation
\be
\beta^{-2l} a_l = \sum_{m,n} \cC^{m n}_l a_m a_n
\ee
is given by
\be
2^{\frac{l}{\lst}}\sum_{k=0}^N (\beta^{-2l})^{(N-k)} a_l^{(k)} =
\sum_{k=0}^N \sum_{m,n} \cC^{m n}_l a_m^{(N-k)} a_n^{(k)} \, .
\ee
Since $a_l^{(0)}=\delta_{l,0}$ and $(\beta^{-0})^{(n)}=0$ for $n>0$, we
obtain
\be
2^{\frac{l}{\lst}}\left( a_l^{(N)} +
 \sum_{k=1}^{N-1} (\beta^{-2l})^{(N-k)} a_l^{(k)} \right) =
 2a_l^{(N)} +
\sum_{k=1}^{N-1} \sum_{m,n} \cC^{m n}_l a_m^{(N-k)} a_n^{(k)} \, .
\label{e2.4}
\ee
We have used $\cC^{m 0}_l = \delta_{m,l}$. Eq.~(\ref{e2.4}) implies for
$l\ne \lst $ eq.~(\ref{e2.2}). Consider eq.~(\ref{e2.4}) for the case $N=1$ :
\be
2^{\frac{l}{\lst}} a_l^{(1)} = 2 a_l^{(1)}.
\ee
This implies $a_l^{(1)} = \alpha \delta_{l,\lst }.$ To determine the
constant $\alpha $, we have to consider eq.~(\ref{e2.4}) for the case $N=2$,
$l=\lst $ :
\be
2 (\beta^{-2\lst})^{(1)} a_{\lst}^{(1)} =  \cC^{\lst  \lst }_l
(a_{\lst}^{(1)})^2 \, .
\ee
This implies
\be
\alpha := \frac{2(\beta^{-2\lst})^{(1)}}{\cC^{\lst \lst}_{\lst}}.
\ee
Let us consider eq.~(\ref{e2.4}) for the case $N\ge 3$, $l=\lst $:
\be
2 \sum_{k=1}^{N-1} (\beta^{-2\lst})^{(N-k)} a_{\lst}^{(k)} =
\sum_{k=1}^{N-1} \sum_{m,n} \cC^{m n}_\lst a_m^{(N-k)} a_n^{(k)}.
\ee
This implies
\begin{eqnarray}
 & & 2 \biggl [ (\beta^{-2\lst})^{(N-1)}\alpha +
(\beta^{-2\lst})^{(1)} a_{\lst}^{(N-1)} +
\sum_{k=2}^{N-2} (\beta^{-2\lst})^{(N-k)} a_{\lst}^{(k)}
\biggr ] =\nonumber\\
&=&
2\sum_m \cC^{m \lst}_\lst a_m^{(N-1)}\alpha +
\sum_{k=2}^{N-2} \sum_{m,n} \cC^{m n}_\lst a_m^{(N-k)} a_n^{(k)}.
\end{eqnarray}
Thus,
\begin{eqnarray}
a_{\lst }^{(N-1)} &=& \frac{1}{2((\beta^{-2\lst})^{(1)}-
\cC^{\lst  \lst }_\lst \alpha )} \biggl [ -2 \biggl (
(\beta^{-2\lst})^{(1)}\alpha +
\sum_{k=2}^{N-2} (\beta^{-2\lst})^{(N-k)} a_{\lst}^{(k)} \biggr )
+\nonumber\\
&+&
2\sum_{m:\ m\ne \lst } \cC^{m \lst}_\lst a_m^{(N-1)}\alpha +
\sum_{k=2}^{N-2} \sum_{m,n} \cC^{m n}_\lst a_m^{(N-k)} a_n^{(k)}
\biggr ].
\end{eqnarray}
Using $\cC_{\lst }^{\lst  \lst} \alpha = 2(\beta^{-2\lst})^{(1)}$
and replacing $N$ by $N+1$, we obtain eq.~(\ref{e2.3}).
$\ \ \Box $

\vskip5mm\noindent
With the following lemma we provide explicit
expressions for the $\epsilon $-expansion of $\beta^{-2l}$.
\begin{lemma}\label{lemma1.1}
Suppose that the coefficients $(\beta^{-l})^{(m)}$ of the
$\epsilon $-expansion for the term $\beta^{-l}$ are defined by
\be
\beta^{-l} = 2^{\frac{l}{2\lst}} \sum_{m=0}^\infty (\beta^{-l})^{(m)}
\epsilon^m \, .
\ee
Then, we have $(\beta^{-l})^{(0)} =1$ and
\be
(\beta^{-l})^{(m)} = \sum_{k=0}^{m-1} \frac{1}{(m-k)!} \left(-
  \frac{l \ln 2}{\dst } \right)^{m-k} {m-1 \choose k} \dst^{-m},
\ee
for $m \ge 1$.
\end{lemma}
\vskip5mm\noindent
\underline{\sl Proof: \,} We have
\be
\beta^{-l} = 2^{\frac{l}{2\lst}} 2^{\frac{l}{\dst }(1- \frac{\dst}{d})} =
    2^{\frac{l}{2\lst }} \exp \left( \frac{l \ln 2}{\dst } \left(1-
    \frac{1}{1-\frac{\epsilon }{\dst }}\right) \right) \, .
\label{e1.1}
\ee
Expansion of the exponential-function on the rhs of eq.~(\ref{e1.1}) gives
\be
\beta^{-l} = 2^{\frac{l}{2\lst }} \left[1+\sum_{n=1}^\infty \frac{1}{n!}
\left( \frac{l \ln 2}{\dst}\right)^n
\left(1-\frac{1}{1-\frac{\epsilon}{d}} \right)^n \right] \, .
\label{e1.2}
\ee
Furthermore
\be
\left(1-\frac{1}{1-\frac{\epsilon }{\dst }} \right)^n =
\left(-\frac{\epsilon}{\dst} \right)^n \sum_{k=0}^\infty {n+k-1 \choose k}
\left(\frac{\epsilon}{\dst} \right)^k.
\label{e1.4}
\ee
Insertion of eq.~(\ref{e1.4}) into eq.~(\ref{e1.2}) yields
\be
\beta^{-l} = 2^{\frac{l}{2\lst}} \left[1+ \sum_{n=1}^\infty \frac{1}{n!}
\left(-\frac{l \ln 2}{\dst} \right)^n \sum_{k=0}^\infty {n+k-1 \choose k}
\left(\frac{\epsilon}{\dst} \right)^{n+k} \right].
\ee
Introducing a new variable $m=n+k \in \{ 1,2,\ldots \} $, we obtain
\be
\beta^{-l} = 2^{\frac{l}{2\lst}} \biggl \{ 1+ \sum_{m=1}^\infty \biggl [
\sum_{k=0}^{m-1} \frac{1}{(m-k)!}
\left(-\frac{l \ln 2}{\dst} \right)^{m-k} {m-1 \choose k}
\dst^{-m}\biggr ] \epsilon^m\biggr \}.
\ee
This implies the assertion. $ \ \ \Box  $

\subsection{$\epsilon$-Expansion for the Linearized RG}
\label{epslin}

We shall now consider the eigenvalue problem for
the linearized RG equation. The eigenvalues and
eigenvectors are computed by using $\epsilon$-expansion.

The linearized RG transformation is given by the
matrix $U(a)$, with matrix elements
\be
U_{nl}(a) := 2\beta^{2l} \sum_m \cC^{m n}_l a_m.
\ee

\begin{lemma}\label{lemma3.5}
Consider the eigenvalue equation for the linearized RG
equation
\be
 U(a)b = \l b \, .
\ee
Suppose that the vector $a$ in $d=\dst -\epsilon $ dimensions, $\dst =
\frac{2\lst}{\lst -1}, \lst \in \{ 2,3,\ldots \} $ is given by the
$\epsilon$-expansion
\be
a_m = \sum_{k:\ k \ge 0} a_m^{(k)} \epsilon^k \, ,
\ee
and the $\epsilon $-expansion of $U$ is
\be
 U=\sum_k U^{(k)} \epsilon^k \, .
\ee
Suppose that the $\epsilon$-expansions of $b$ and $\l $ are given by
\be
 b = \sum_{m:\  m\ge 0} b^{(m)} \epsilon^m, \qquad \l  =
 \sum_{n:\  n\ge 0} \l^{(n)} \epsilon^n .
\ee
Suppose that $b_m^{(0)} =  \delta_{n,\om }$
and $\l^{(0)} = 2^{1-\frac{\om}{\lst}}$, for $\om \in \{ 0,1,2,\dots \} $.
Then, we have
\be
\l^{(N)} = \sum_{n=0}^{N-1} \sum_m
            U_{\om m}^{(N-n)} b_m^{(n)} \, ,
\label{e3.20}
\ee
and, for $l\ne \om$,
\be
b_l^{(N)} = \frac{1}{2(2^{-\frac{l}{\lst}}-2^{-\frac{\om}{\lst}})}
    \sum_{n=0}^{N-1} \sum_m \left(\l^{(N-n)}\delta_{l,m} -
      U_{lm}^{(N-n)} \right) \, b_m^{(n)} \, .
\label{e3.21}
\ee
Furthermore,
\be
b_{\om}^{(N)} = 0 \, .
\ee
\end{lemma}
The proof of this lemma will be given later.
\begin{lemma}\label{lemma3.1}
$U(a)$ is symmetric with respect to the canonical scalar product,
defined by
\be
(a,b):= \sum_n  a_n b_n \, .
\label{e3.1}
\ee
\end{lemma}
\vskip5mm\noindent
\underline{\sl Proof:\,} We have
\be
(u,U(a)v) = \sum_{n,l} u_n T_{nl} v_l,
\ee
where
\be
T_{nl} :=  \frac{\beta^{2n}}{(2n)!} U_{nl}
\ee
Since $T_{nl} = T_{ln}$, we have
\be
(u,U(a)v) = (U(a)u,v). \ \ \Box
\ee
Lemma \ref{lemma3.1} shows that the
eigenvalues of $U(a)$ are real.
The next lemma shows how to compute the $\epsilon$-expansion for
the eigenvectors and eigenvalues
\begin{lemma}\label{lemma3.2}
Suppose that the linearized RG group equation
is given by the following series expansion
\be
U(a) = \sum_{n:\ n\ge 0} U^{(n)}(a) \epsilon^n
\ee
and that the 0th order term $U^{(0)}$ is symmetric.
Let $b$ be an eigenvector with eigenvalue $\l $ of $U(a)$, i.~e.
\be
U(a)b=\l b.
\ee
Let $b^{(0)}$ be a normalized eigenvector with eigenvalue $\l^{(0)}$
of $U^{(0)}(a)$, i.~e.
\be
U^{(0)}(a)b^{(0)}=\l^{(0)} b^{(0)}, \qquad (b^{(0)},b^{(0)})=1,
\ee
where the scalar product $(\ ,\ )$ is defined by eq.~(\ref{e3.1}).
Suppose that the eigenvalue $\l^{(0)}$ is not degenerate. Then,
there exists an $\epsilon$-expansion for the eigenvector $b$ with
eigenvalue $\l $
\be
b=\sum_m b^{(m)}\epsilon^m, \qquad \l = \sum_n \l^{(n)}\epsilon^n
\ee
such that
\be
(b^{(0)},b^{(N)})= 0,
\label{e3.9}
\ee
for all $N>0$. The coefficients $b^{(N)}$ and $\l^{(N)}$
are recursively determined by
\be
\l^{(N)} = \sum_{n=0}^{N-1} (b^{(0)}, U^{(N-n)} b^{(n)})
\label{e3.10}
\ee
and
\be
b^{(N)} = (U^{(0)} - \l^{(0)})^{-1}
\left[\sum_{n=0}^{N-1} (U^{(N-n)}-\l^{(N-n)}) b^{(n)} \right]^\bot ,
\label{e3.11}
\ee
where $u^\bot $ is the component of $u$ perpendicular to $b^{(0)}$.
$\ \ \Box $
\end{lemma}
\vskip5mm\noindent
\underline{\sl Proof:\,} The eigenvalue equation implies,
for all $N$,
\be
\sum_{m,n: m+n=N} (U^{(m)} b^{(n)} - \l^{(m)}b^{(n)}) =0.
\ee
Thus,
\be
(U^{(0)} - \l^{(0)})b^{(N)} =
\sum_{n=0}^{N-1} (U^{(N-n)} - \l^{(N-n)}) b^{(n)}.
\label{e3.12}
\ee
Since $(U^{(0)} - \l^{(0)})b^{(0)} = 0$,
we may add to $b^{(N)}$, $N>0$ on the
lhs of eq.~(\ref{e3.12}) a multiple of vector $b^{(0)}$
such that eq.~(\ref{e3.9}) holds. Scalar
multiplication of eq.~(\ref{e3.12}) with $b^{(0)}$ gives
\be
(b^{(0)},(U^{(0)} - \l^{(0)})b^{(N)}) =
\sum_{n=0}^{N-1} (b^{(0)},(U^{(N-n)} - \l^{(N-n)}) b^{(n)}).
\label{e3.13}
\ee
Since $U^{(0)}$ is symmetric, the lhs of eq.~(\ref{e3.13}) is zero.
Therefore eq.~(\ref{e3.10}) is valid.
Since the rhs of eq.~(\ref{e3.12}) is perpendicular
to $b^{(0)}$, eq.~(\ref{e3.11}) can be computed by eq.~(\ref{e3.12}).
$\  \  \Box $
\par\medskip
The next lemma presents the $\epsilon $-expansion of the linearized
RG transformation $U(a)$.
\begin{lemma}\label{lemma3.3}
Let the $\epsilon $-expansion of $\beta^{2l}$ be
\be
\beta^{2l} = 2^{-\frac{l}{\lst}} \sum_{m=0}^\infty (\beta^{2l})^{(m)}
  \epsilon^m, \qquad (\beta^{2l})^{(0)}=1,
\ee
for $\lst \in \{ 2,3,\ldots \} $.
Then the $\epsilon $-expansion of $U$
\be
U=\sum_k U^{(k)} \epsilon^k
\ee
is explicitly given by
\be
U_{nl}^{(k)} = 2^{1-\frac{l}{\lst}} \sum_{r=0}^k \sum_m
\cC^{m n}_l (\beta^{2l})^{(k-r)} a_m^{(r)}.
\ee
\end{lemma}
\vskip5mm\noindent
The proof of the foregoing lemma follows immediately from the definitions.
\par\medskip
For the recursive computation of the coefficients for the eigenvalues
and eigenvectors we need the start values of the recursion relations
which are
eigenvalues and eigenvectors of $U^{(0)}$.
\begin{lemma}\label{lemma3.4}
Suppose that $a_m^{(0)} = \delta_{m,0}$. The normalized eigenvectors
$b^{(0)}$ with eigenvalues $\l^{(0)}$ of $U^{(0)}$ are
\be
b^{(0)}_m = \delta_{m,\om }, \qquad
\l^{(0)} = 2^{1-\frac{\om}{\lst}} ,
\ee
for all $\om \in \{ 0,1,2,\ldots \} .$
\end{lemma}
\vskip5mm\noindent
\underline{\sl Proof:\,}
We have, using $a_m^{(0)} =\delta_{m,0}$, $\cC^{0 n}_l = \delta_{l,n}$
\be
U_{nl}^{(0)} = 2^{1-\frac{l}{\lst}} \delta_{n,l}.
\ee
Thus
\be
(U^{(0)} b^{(0)})_n = 2^{1-\frac{\om}{\lst}}
        \delta_{n,\om } =
            2^{1-\frac{\om}{\lst}} b^{(0)}_n. \  \  \Box
\ee
\vskip1cm\noindent
We finish this subsection with the proof of \ref{lemma3.5}.

\noindent
\underline{\sl Proof of Lemma \ref{lemma3.5}:} By Lemma \ref{lemma3.2},
eq.~(\ref{e3.10}), follows eq.~(\ref{e3.20}).
Since
\be
U_{nl}^{(0)} = 2^{1-\frac{l}{\lst}}\delta_{n,l}, \qquad
   \l^{(0)} =  2^{1-\frac{\om}{\lst}} \, ,
\ee
we have
\be
(U^{(0)} -\l^{(0)})_{nl} = 2(2^{-\frac{l}{\lst}}-2^{-\frac{\om}{\lst}})
          \delta_{n,l} \, .
\ee
Therefore
\be
(U^{(0)} -\l^{(0)})^{-1}_{nl} =
  \frac{1}{2(2^{-\frac{l}{\lst}}-2^{-\frac{\om}{\lst}})}
          \delta_{n,l}.
\ee
Thus, eq.~(\ref{e3.11}) of Lemma \ref{lemma3.2} implies
eq.~(\ref{e3.21}). $\ \ \Box $

\subsection{Comparison of $\epsilon$-Expansion with Numerical Results}
\label{comparison}

We evaluated the recursion relations presented in the preceding
subsection using the computer algebra program Maple V Release 2.
This allowed us to go to relatively high order. We always used
programs that computed everything exactly (in the form of analytical
expressions) and programs that solved the recursion relations
numerically. Note however, that Maple allows for arbritrary high
precision in the numerical computations.
It was no problem to compute the coefficients $a_l^{(n)}$ for
$d^*=4$ exactly to sixth order in $\epsilon$. However, the expressions
become quite nasty then. As an example we present the coefficients
$a_l^{(n)}$ up to $n=3$ in Appendix 2. The general structure for
the expansion around $d^*=4$ is that
at a given order $n$ the only nonvanishing coefficients are those
with $l \leq 2 n$. The corresponding relation at $d^*=3$ is
$l \leq 3 n$.
The $\epsilon$-expansion is expected to have zero convergence
radius. However, the series are believed to be Borel-summable.
For small $\epsilon$ even the naively summed
low order series can be a quite good approximation.
In figure 4 we show the comparison of the `true' 2-well potential
at $d=3.8$ with the 1st and 4th order $\epsilon$-expansion.
The full line gives the result obtained numerically, and the
dashed lines give the 1st and 4th order approximations, respectively.

With the numerical version of the program it was no problem to go
to orders like 16.
In table \ref{growthtab} we show the intimidating growth of
the expansion coefficients when the order becomes large.

\begin{table}
\begin{center}
\begin{tabular}{|c|r|r|}
\hline
  \mc{1}{|c|}{$n$}  & \mc{1}{c|}{$a_1^{(n)}$} & \mc{1}{c|}{$a_2^{(n)}$}  \\
\hline
  $10$ & $ 4.59314 \cdot 10^{1}$ & $ -2.89444 \cdot 10^{2}$ \\
  $11$ & $-3.41664 \cdot 10^{2}$ & $  2.26283 \cdot 10^{3}$ \\
  $12$ & $ 2.70284 \cdot 10^{3}$ & $ -1.87364 \cdot 10^{4}$ \\
  $13$ & $-2.26138 \cdot 10^{4}$ & $  1.63588 \cdot 10^{5}$ \\
  $14$ & $ 1.99286 \cdot 10^{5}$ & $ -1.50104 \cdot 10^{6}$ \\
  $15$ & $-1.84404 \cdot 10^{6}$ & $  1.44371 \cdot 10^{7}$ \\
  $16$ & $ 1.78725 \cdot 10^{7}$ & $ -1.45243 \cdot 10^{8}$ \\
\hline
 \end{tabular}
  \parbox[t]{.85\textwidth}
  {
  \caption[nui]{\label{growthtab}
  Numerical results of $a_1$ and $a_2$ for $n$ from $10$ to $16$}
  }
\end{center}
\end{table}
With the help of the recursion relations of the preceding
subsections we also determined the $\epsilon$-expansion for
the exponent $\nu$. We again used an exact version of the
program that was practicable up to order 6, and a numerical
version that could be used to higher order.

In table \ref{tabnui} we show our results for the expansion
coefficients of $\nu$, compared with those of the
full model \cite{BuchZinnJustin}. The first two orders
are exactly the same (the coefficient $\nu_1$ is $1/12$
in the hierarchical and in the full model). This might
be due to the fact that the $\epsilon$-expansion of
the exponent $\eta$ starts at order $\epsilon^2$.

\begin{table}
\begin{center}
\begin{tabular}{|r|r|r|}
\hline
  \mc{1}{|c|}{$i$}  & \mc{1}{c|}{$\nu_{i}$} & \mc{1}{c|}{$\nu_{i,f}$}  \\
\hline
     0 &     0.5000 &     0.5000  \\
     1 &     0.0833 &     0.0833  \\
     2 &     0.0556 &     0.0445  \\
     3 &    -0.0324 &    -0.0190  \\
     4 &     0.1468 &     0.0888  \\
     5 &    -0.5743 &    -0.2015  \\
\hline
 \end{tabular}
  \parbox[t]{.85\textwidth}
  {
  \caption[nui]{\label{tabnui}
  Comparison of the $\epsilon$-expansion coefficients of $\nu$ in
  the hierarchical model $\nu_i$ with the ones in the full model
  $\nu_{i,f}$}
  }
\end{center}
\end{table}
In table \ref{tabpad} we give the results of resummed series
for $\nu$ up to order $\epsilon^k$, for $k \leq 5$, and
$d=3.0 \dots3.8 $. For the larger values of $\epsilon$,
the signal of the divergence of the series is obvious.
For comparison we also quote our numerical result (`true')
and the result of a Borel-Pad\`e summation of the
6th order $\epsilon$ expansion (BP). The latter was obtained
as follows: The $\nu_i$ were divided by $i!$, and the
diagonal Pad\`e approximation of the resulting Taylor series
was determined. From the resulting rational function $Q(\epsilon)$ the
estimate for $\nu$ was then obtained by numerically computing
\be
\nu(\epsilon) = \int_{0}^{\infty} dt \, \exp(-t) \, Q(t \epsilon) \, .
\ee
There is a quite good agreement with the `true' results.
(For $d=3$, the diagonal Pad\'e of the Borel transform had a
nonintegrable singularity on the positive real axis.)

\begin{table}
\begin{center}
\begin{tabular}{|c|c|c|c|c|c|c|c|}
\hline
$d$  & $k=1$ & $k=2$ & $k=3$ & $k=4$ & $k=5$ & `true' & BP      \\
\hline
 3.0 & 0.6161 & 0.6523 & 0.6060 & 0.8255 & 0.3639 & 0.64957 &         \\
 3.1 & 0.6007 & 0.6293 & 0.5961 & 0.7228 & 0.4235 & 0.62570 & 0.62599  \\
 3.2 & 0.5864 & 0.6084 & 0.5855 & 0.6580 & 0.4771 & 0.60484 & 0.60136  \\
 3.3 & 0.5730 & 0.5895 & 0.5744 & 0.6144 & 0.5160 & 0.58640 & 0.58791  \\
 3.4 & 0.5605 & 0.5723 & 0.5630 & 0.5837 & 0.5369 & 0.56995 & 0.56973  \\
 3.5 & 0.5487 & 0.5568 & 0.5515 & 0.5612 & 0.5423 & 0.55516 & 0.55525  \\
 3.6 & 0.5377 & 0.5428 & 0.5402 & 0.5440 & 0.5379 & 0.54181 & 0.54183  \\
 3.7 & 0.5274 & 0.5302 & 0.5291 & 0.5303 & 0.5289 & 0.52973 & 0.52973  \\
 3.8 & 0.5177 & 0.5190 & 0.5186 & 0.5189 & 0.5187 & 0.51877 & 0.51877  \\
\hline
 \end{tabular}
  \parbox[t]{.85\textwidth}
  {
  \caption[Naive resummation of the nu series]{\label{tabpad}
  Results of resummed series
  for $\nu$ up to order $\epsilon^k$, for $k \leq 5$.
  `true' is the numerical result, BP is the result of a
  Borel-Pad\'e summation explained in the text}
  }
\end{center}
\end{table}
\section{Conclusions}

In this paper, we have demonstrated that at least for hierarchical
models, an algebraic computation of fixed points and exponents  is
feasible. An extension to $N$-component models and general values of $L$
 could be easily done. Of course, many new ideas are necessary to  do
the same thing in full models. The crucial question here  is the proper
choice of parametrization of the Boltzmannian. An interesting question
which certainly deserves study  is whether the $\epsilon$-expansion for
the RG flow   (in the Wilson sense) in full models could be performed
to an order that is competitive with what has been done  in the
conventional framework.

\section*{Acknowledgment}
We are grateful to Peter Wittwer for sharing his insights with
us. We also thank him for the possibility to compare some of our
numerical results with his.
A.P. and C.W. would like to thank the Deutsche Forschungsgemeinschaft
for financial support under grant Wi~1280/2-1.

\section*{Appendix 1: Field Theory and Hierarchical Models}
This appendix  discusses the hierarchical model as an approximation
of a field theoretic model and motivates the hierarchical RG equation.

The generating functional for Greens functions of a scalar field theory on a
$d$-di\-men\-si\-o\-nal continuum $\bR^d$ is given by the following (formal)
infinite-dimensional integral
\be
Z[J] := \cN \int \prod_{z\in \bR^d} d\phi (z) \,
        \exp \left[ -\half (\phi , -\Delta \phi ) \right ]
        \exp \left[ -V(\phi)+(J,\phi )\right] \, .
\ee

$\phi $ is a scalar field, and $J$ is an external source on $\bR^d$.
$V(\phi ) = \int_{z\in \bR^d}
\cV \bigl (\phi (z)\bigr ) $
is the local interaction term, and $\frac{1}{2}(\phi , -\Delta \phi )$
is the free part of the action. $\Delta$ is the Laplacian, and
$\cN $ is a normalization factor chosen such that
\be
\Gau{\rv}{\phi} := \cN \prod_{z\in \bR^d} d\phi (z) \,
        \exp \left[ -\half (\phi , -\Delta \phi )\right]
\ee
is a normalized Gaussian measure. $\rv$ is called propagator.
The canonical scalar product is defined by
\be
(\phi ,\psi ) := \int_{z\in \bR^d} \phi (z) \psi (z),
\ee
for fields $\phi ,\psi $ on $\bR^d$.

For RG calculations it is more convenient to use
generating functionals with external fields $\psi $,
\be \label{gener}
Z(\psi ):= \iGaurv \, \exp \left[ -V(\phi +\psi ) \right] \, .
\ee
The two generating functionals are related by
\be
Z(\psi )= Z[J] \, \exp \left[ -\half (J,\rv J) \right]
                        \biggr\vert_{J=\rv^{-1}\psi} \, .
\ee
For the definition of the hierarchical model, we introduce the notion of
a  hierarchical lattice (or multigrid). For $L\in \{ 2,3,\dots \} $ and
$j\in \bZ $ divide $\bR^d$ into hypercubes of side length $a_j:=
L^{-j}a$, where $a$ is a unit length. Denote the set of all  these
hypercubes by $\L_j$. $\L_j$ can be considered as a lattice with lattice
spacing $a_j$ by identifying the centers of the hypercubes with the
lattice sites. The location of the hypercubes can be chosen in the
following way. For a hypercube $y\in \L_j$ let  $\tilde{y}$ be the open
hypercube of $y$. For $y\in \L_j$ and $x\in \L_k$ we suppose that
$\tilde{y} \cap x =\emptyset $ or $\tilde{y} \subseteq x$. In the latter
case we write $y\inn x$. The hierarchical approximation is given by the
following replacement
\be \label{hierep}
\rv  \longrightarrow \sum_{j\in \bZ } \rv^j,
\ee
where, for $z,\zp \in \bR^d$,
\be \label{hiefluc}
\rv^j(z,\zp ) := a_j^{2-d} \, \g  \, \delta_{x,\xp } \, .
\ee
Here, $x$ and $\xp $ are the uniquely determined hypercubes of $\L_j$
such that $z\in x$, $\zp \in \xp $.
$\rv^j$ is called fluctuation propagator.
The above replacement represents
the fact that in general $\rv $ can be decomposed into a sum of
fluctuation propagators
which are exponentially decaying with decay length $a_j$. This exponential
decay is simulated by the Kronecker delta on the rhs
of eq.~(\ref{hiefluc}).
There are other ways to define the hierarchical approximation. But all
hierarchical approximations share the property that the fluctuation
propagators $\rv^j$ have compact support.
Insertion of replacement (\ref{hierep}) into eq.~(\ref{gener}) yields,
using the convolution formula of Gaussian measures,
\be
Z(\psi )= \int \prod_{j\in \bZ } d\mu_{\rv^j} (\phi^j )
  \exp \biggl[ -V(\sum_{j\in \bZ } \phi^j +\psi ) \biggr] \, .
\ee
We define an ultraviolet cutoff by setting the propagator $\rv^j =0$
if $j>n$ and an infrared cutoff by setting $\rv^k=0$ if $k\le j$ :
\be
Z_j^{(n)}(\psi ) = \int \prod_{i=j+1}^n d\mu_{\rv^i} (\phi^i )
  \exp \biggl[ -V(\sum_{i=j+1}^n \phi^i +\psi ) \biggr] \, .
\ee
The effective generating functions $Z_j^{(n)}$ and $Z_{j-1}^{(n)}$
are related by
\be \label{renfull}
Z_{j-1}^{(n)}(\psi ) = \int d\mu_{\rv^j} (\phi^j ) Z_j^{(n)} (\phi^j +
  \psi ) \, .
\ee
Since we started with a local interaction $V$, the effective generating
functions obey the following factorization property
\be
Z_j^{(n)}(\psi ) = \prod_{y:\, y\in \L_j} Z_j^{(n)}(y|\psi ) \, .
\ee
Since the kernel of $\rv^j$ is constant on hypercubes of $\L_j$, we can
assume that the fields $\psi $ are constant arguments of $Z_j^{(n)}$,
i.e., do not depend on $z \in \bR^d$.
Therefore, eq.~(\ref{renfull}) is equivalent to
\be \label{renfac}
Z_{j-1}^{(n)}(x|\psi ) = \prod_{y:\, y\inn x}
   \biggl [ \int d\mu_{\rv^j} (\phi ) Z_j^{(n)} (y|\phi +
  \psi ) \biggr ],
\ee
for $x\in \L_{j-1}$.
For translation invariant models $Z_j^{(n)} (y|\phi )$ does not depend on
$y\in \L_j$. Let us define
\be \label{zdef}
Z_j^{(n)}(\phi ) := Z_j^{(n)} (y|a_j^{1-\frac{d}{2}}\phi ) \, .
\ee
{}From eq.~(\ref{renfac}) and definition eq.~(\ref{zdef}) follows,
using $\psi \rightarrow a_{j-1}^{1-\frac{d}{2}}\psi $,
\be
Z_{j-1}^{(n)}(\psi ) =
   \biggl [ \int d\mu_{\g } (\phi ) Z_j^{(n)} (\phi + L^{1-\frac{d}{2}}
  \psi ) \biggr ]^{L^d} \, .
\ee
For $F^{'} := (Z_{j-1}^{(n)})^{L^{-d}}$ and $F := (Z_{j}^{(n)})^{L^{-d}}$,
we obtain the hierarchical RG transformation
eq.\ (\ref{trafofull}).
\section*{Appendix 2: Some Results of $\epsilon$-Expansion for the $a_l^{(n)}$}

\begin{eqnarray}
R &:=& \ln 2 \nonumber \\
T &:=& \sqrt{2} \nonumber \\
 &\quad& \\
a_   0   ^{(   0   )}   & =&
1
\nonumber \\
a_   0   ^{(   2   )}   & =&
-{\frac {R^{2}}{864}}
\nonumber \\
a_   0   ^{(   3   )}   & =&
{\frac {12\,R^{2}\left (3\,R-2\right )-R^{2}\left (19\,R-18\right )T}{
-31104\,T+41472}}
\nonumber \\
a_   1   ^{(   2   )}   & =&
{\frac {R^{2}}{216\,T-432}}
\nonumber \\
a_   1   ^{(   3   )}   & =&
{\frac {-2\,R^{2}\left (7\,R-8\right )+R^{2}\left (7\,R-12\right )T}{
17280\,T-24192}}
\nonumber \\
a_   2   ^{(   1   )}   & =&
-{\frac {R}{144}}
\nonumber \\
a_   2   ^{(   2   )}   & =&
{\frac {12\,R\left (3\,R-2\right )-R\left (19\,R-18\right )T}{-10368\,
T+13824}}
\nonumber \\
a_   2   ^{(   3   )}   & =&
{\frac {R\left (154\,R^{2}+1161\,R-459\right )-12\,R\left (8\,R^{2}+69
\,R-27\right )T}{-746496\,T+1057536}}
\nonumber \\
a_   3   ^{(   2   )}   & =&
{\frac {R^{2}}{2592\,T-2592}}
\nonumber \\
a_   3   ^{(   3   )}   & =&
{\frac {-8\,R^{2}\left (R-1\right )+R^{2}\left (5\,R-6\right )T}{72576
\,T-103680}}
\nonumber \\
a_   4   ^{(   2   )}   & =&
{\frac {R^{2}}{41472}}
\nonumber \\
a_   4   ^{(   3   )}   & =&
{\frac {-12\,R^{2}\left (3\,R-2\right )+R^{2}\left (35\,R-18\right )T}
{-1492992\,T+1990656}}
\nonumber \\
a_   5   ^{(   3   )}   & =&
-{\frac {R^{3}}{373248\,T-373248}}
\nonumber \\
a_   6   ^{(   3   )}   & =&
-{\frac {R^{3}}{17915904}}
\nonumber \\
\end{eqnarray}


\newpage

\vskip3cm
\noindent
{\Large \bf Figure Captions}

\vskip1cm
\noindent
{\bf Fig.\ 1:}
Our results for the 2-well potential  for $d=2.1$ through $d=3.8$ in
steps of $0.1$. The deepest potential  corresponds to $d=2.1$.  With
increasing dimension, the 2-well  gets flatter and flatter until it
vanishes in 4 dimension. In all cases we used $\lmax=20$.

\vskip3mm\noindent
{\bf Fig.\ 2:}
The fixed point 3-wells for $d=2.1$ through $d=2.7$ in steps of $0.1$.
The deepest potential corresponds to $d=2.1$, the flattest
one to $d=2.7$. In all cases we used $\lmax=20$.

\vskip3mm\noindent
{\bf Fig.\ 3:}
The fixed point 4-well potentials for $d=2.1$ and $d=2.3$.
The potential with the deeper wells corresponds to $d=2.1$.
In both cases we used $\lmax=30$.

\vskip3mm\noindent
{\bf Fig.\ 4:}
Comparison of the `true' 2-well potential
at $d=3.8$ with the 1st and 4th order $\epsilon$-expansion.
The full line gives the result obtained numerically, and the
dashed lines give the 1st and 4th order approximations, respectively.

\vskip3mm\noindent
{\bf Fig.\ 5:}
Comparison of the two leading eigenvectors of the RG transformation
linearized around the trivial fixed point and the 2-well fixed point
in $d=3$. The single well shaped functions  correspond to the
relevant eigenvalue. The double well shaped funtions correspond
to the first irrelevant eigenvalue. The full lines belong to
the non-trivial fixed point.
\end{document}